\documentclass{emulateapj}

\usepackage{amsmath}
\usepackage{natbib}
\usepackage{float}
\usepackage{dcolumn}
\usepackage{booktabs}
\usepackage[
	colorlinks=true,
	urlcolor=blue,
	linkcolor=blue,
	citecolor=blue
	]{hyperref}
\usepackage{rotating}
\usepackage{graphicx}
\usepackage{tablefootnote}
\usepackage{color}

\slugcomment{Accepted by the Astrophysical Journal}

\shorttitle{Main-sequence stars masquerading as Young Stellar Objects in the central molecular zone}
\shortauthors{Koepferl, Robitaille, Morales, Johnston}

\begin{document}

\title{Main-sequence stars masquerading as Young Stellar Objects in the central molecular zone}
\author{Christine M. Koepferl, Thomas P. Robitaille, Esteban F. E. Morales, Katharine G. Johnston}
\affil{Max Planck Institute for Astronomy, K\"onigstuhl 17, 69117 Heidelberg, Germany}
\journalinfo{Accepted by the Astrophysical Journal}
\submitted{Received 2014 July 13; accepted 2014 November 13}
\email{koepferl@mpia.de}

\begin{abstract}
In contrast to most other galaxies, star-formation rates in the Milky Way can be estimated directly from Young Stellar Objects (YSOs). In the Central Molecular Zone (CMZ) the star-formation rate calculated from the number of YSOs with 24\,$\mu$m emission is up to order of magnitude higher than the value estimated from  methods based on diffuse emission (such as free-free emission). Whether this effect is real or whether it indicates problems with either or both star formation rate measures is not currently known. In this paper, we investigate whether estimates based on YSOs could be heavily contaminated by more evolved objects such as main-sequence stars. We present radiative transfer models of YSOs and of main-sequence stars in a constant ambient medium which show that the main-sequence objects can indeed mimic YSOs at 24\,$\mu$m. However, we show that in some cases the main-sequence models can be marginally resolved at 24\,$\mu$m, whereas the YSO models are always unresolved. Based on the fraction of resolved MIPS 24\,$\mu$m sources in the sample of YSOs previously used to compute the star formation rate, we estimate the fraction of misclassified ``YSOs" to be at least 63\,\%, which suggests that the star-formation rate previously determined from YSOs is likely to be at least a factor of three too high.
\end{abstract}

\section{Introduction}
\label{intro}
Various indirect techniques based on diffuse emission are traditionally applied in order to measure the star formation rate (SFR) of galaxies throughout the Universe. These common techniques use star-formation tracers such as free-free cm continuum, H$\alpha$, and far infrared emission to indirectly infer the rate of forming stars (for details see review by \citealp{2013:Calzetti}). Free-free emission (or Bremsstrahlung) emitted from ionized electrons, and recombination lines of ionized hydrogen (e.\,g.~Balmer series H$\alpha$) both trace gas ionized by young massive stars. Another method is to use the infrared flux as a tracer, since the UV radiation from young massive stars is also absorbed by surrounding dust and re-emitted in the infrared. Combinations of various diffuse tracers (such as 24\,$\mu$m emission together with H$\alpha$) are now commonly used. All these conventional methods usually agree but trace only the high-mass star formation rate, and require an extrapolation to lower mass stars. 
 
For the Milky-Way, we have the opportunity to directly estimate the SFR by counting individual young or forming stars, which if calibrated with caution is a preferred method as it accounts for the actual sites of star-formation. With the \emph{Spitzer} mid-infrared survey "cores-to-disks" (c2d, \citealp{2009:Evans}) the SFR of nearby star-forming regions was estimated to be 6.5$\times 10^{-6}$ to 9.6$\times 10^{-5}$\,$\mbox{M}_{\odot}\,\mbox{yr}^{-1}$ by counting YSOs (down to low mass objects), assuming an average mass of 0.5\,$\mbox{M}_\odot$ and a star-formation duration of 2\,Myr. For the first time \cite{2010:Robitaille} calculated the total SFR of Milky Way Galaxy by counting sources showing a mid-infrared excess at IRAC wavelengths, using a population synthesis model to extrapolate the number of sources beyond the detection limit. They found an overall SFR for the Milky Way of 0.68 to 1.45\,$\mbox{M}_{\odot}\,\mbox{yr}^{-1}$. This is in agreement with techniques based on diffuse emission tracers: indeed, \cite{2011:Chomiuk} demonstrated that once the SFRs derived from the different methods are normalized to the same assumptions (for example for the IMF), the methods are all consistent with a value of 1.9\,$\pm\,0.4\,\mbox{M}_{\odot}\,\mbox{yr}^{-1}$.

\cite{2009:Yusef} calculated the SFR of the Central Molecular Zone (CMZ) directly from their 599 classified YSOs with 24\,$\mu$m emission and found a rate of 0.14\,$\mbox{M}_{\odot}\,\mbox{yr}^{-1}$. In order to derive this SFR, they determined stellar masses for each YSO, constructed a mass function, and fit a Kroupa Initial Mass function (IMF, \citealp{2001:Kroupa}) to the peak of the derived mass distribution in order to extrapolate the stellar mass below their sensitivity limit, then assumed the upper limit to the age of 1\,Myr for 213 of these objects. When comparing their derived SFR with values from indirect methods (e.\,g.~free-free emission) large differences arise, up to a factor of 10, in contrast with the good agreement seen on the scale of the Milky-Way as a whole. For example, \cite{2013:Longmore} found, from their measurements of the total ionizing flux using the free-free emission, that the overall SFR of the CMZ is roughly 0.012-0.018\,$\mbox{M}_{\odot}\,\mbox{yr}^{-1}$ for $|b|\leq 0.5^\circ$ and 0.06\,$\mbox{M}_{\odot}\,\mbox{yr}^{-1}$ for $|b|\leq 1^\circ$, also covering areas outside the CMZ. 

\begin{figure*}
\includegraphics[width=0.97\textwidth]{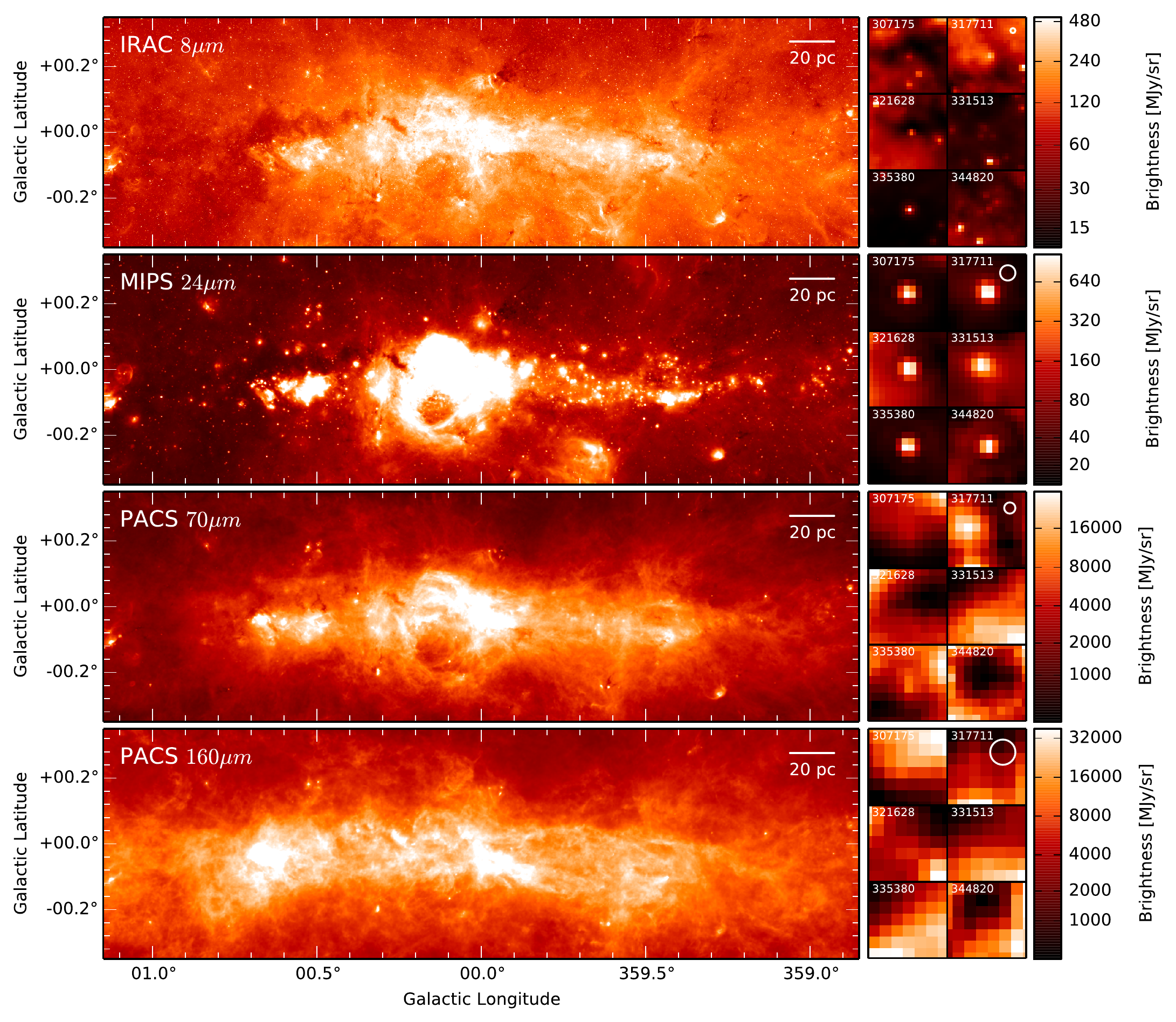}
\caption{IRAC 8\,$\mu$m, MIPS 24\,$\mu$m, PACS 70\,$\mu$m and PACS 160\,$\mu$m observations of the Central Molecular Zone (CMZ) on a arcsinh scale. The active star-forming region Sgr B2 ($\ell\simeq 0.5^\circ$, $b\simeq -0.05^\circ$) and Sgr C ($\ell\simeq 359.4^\circ$, $b\simeq -0.1^\circ$) are very bright in MIPS 24\,$\mu$m and PACS 70\,$\mu$m. The objects in the panels to the right show 30\,x\,30\arcsec\ zoom-ins of the objects classified as YSOs by \cite{2009:Yusef}, which have no strong counterparts in PACS 70\,$\mu$m. The numbers refer to the IDs given by \cite{2009:Yusef}. The color maps of the smaller panels on the right are, in contrast to the main panels, linear and normalized to each panel's individual extrema. The PSF FWHMs are shown in the top right zoom panel for each wavelength.}
\label{Milky Way}
\end{figure*}

Whether the differences are real or due to issues with either the YSO-counting or the diffuse emission methods has not been established. The estimates based on YSOs could be heavily contaminated by reddened objects older than 1\,Myr. In addition to these observational effects, the cm continuum and YSO counting methods may trace different time-scales and thus would be expected to disagree, if the star-formation was episodic. In this paper, we aim to determine how much observational effects are able to contribute to this difference.

To distinguish embedded YSOs from more evolved objects we applied the Stage formalism introduced by \cite{2006:Robitaille}. With this formalism, true YSOs can be separated from more evolved objects. In this formalism, {\bf Stage 0/1} YSOs are very young and envelope dominated, with $\dot{\mbox{M}}_{gas}/\mbox{M}_\star>10^{-6}\,\mbox{yr}^{-1}$. Assuming that the envelope infall rate goes down in time, this corresponds to an upper limit of the timescale of 1\,Myr. Hence, since \cite{2009:Yusef} assumed that their YSOs were less than 1\,Myr old, these would be classified as {\bf Stage 0/1}. In contrast, we group more evolved YSOs and main-sequence stars into the {\bf Stage 2+} category, with $\dot{\mbox{M}}_{gas}/\mbox{M}_\star<10^{-6}\,\mbox{yr}^{-1}$.

In Figure~\ref{Milky Way}, we show the CMZ as observed by the \emph{Spitzer} Space Telescope and the \emph{Herschel} Space Observatory: the top panel shows the 8\,$\mu$m image observed with \emph{Spitzer's} IRAC camera as a part of the GLIMPSE survey \citep{2003:Benjamin, 2009:Churchwell}; below is the \emph{Spitzer} MIPSGAL survey \citep{2009:Carey} using the 24\,$\mu$m band of the MIPS detector; the two lower panels are far-infrared \emph{Herschel} images (70 and 160\,$\mu$m) from the Hi-GAL survey \citep{2010:Molinari} observed with the PACS detector. The \emph{Spitzer} Space Telescope infrared detectors IRAC (3.6, 4.5, 5.8 and 8.0\,$\mu$m) and MIPS (24, 70 and 160\,$\mu$m) have a Point-Spread-Function (PSF) with full-width 1.66, 1.72, 1.88, 1.98\,\arcsec\ and 6, 18, 40\,\arcsec, respectively. By comparison, \emph{Herschel's} PACS detecter at 70, 100 and 160\,$\mu$m has a PSF with full-width of about 4.4, 6.1 and 9.9\,\arcsec. Hereafter, we will refer to the four bands shown in Figure~\ref{Milky Way} as IRAC 8\,$\mu$m, MIPS 24\,$\mu$m, PACS 70\,$\mu$m and PACS 160\,$\mu$m, respectively.

By examining these observations, we found that some of the YSOs classified by \cite{2009:Yusef} appear to have no or weak counterparts in PACS observations (see zoom-in panels in Figure~\ref{Milky Way}), which is counter-intuitive for YSOs. Most of these objects are located up to a Galactic longitude of about $\ell\simeq 359.5^\circ$, to the west of the Galactic center. In contrast, when looking at an active star-forming region (e.\,g.~Sgr B2 at $\ell\simeq 0.5^\circ$) the star formation region as a whole is clearly seen in PACS. 

Obscured main-sequence stars can mimic YSOs, since the surrounding ambient dust is remitting the stellar flux in the infrared (e.g. \citealt{2013:Whitney}). Therefore, in this paper we set out to explore whether objects not seen at wavelength longer than 24\,$\mu$m may not be as young as 1\,Myr (as assumed by \citealt{2009:Yusef}), and hence may not be members of the {\bf Stage 0/1} classification. However, we find instead that although non-detection at PACS wavelengths does not indicate whether a source is a YSO or not, its size at 24\,$\mu$m can be an age indicator.

To determine these results, we set up radiative transfer models and computed realistic synthetic observations of YSOs ({\bf Stage 0/1}) and more evolved objects ({\bf Stage 2+}) in different dust and density environments (Section~\ref{model}). To determine, whether more evolved objects ({\bf Stage 2+}) could mimic YSOs ({\bf Stage 0/1}), we compare the realistic synthetic observations directly with the observations and develop selection criteria that can help reduce contamination from evolved objects such as main-sequence stars (Section~\ref{results}). In Section~\ref{discuss} we discuss the effects of masquerading main-sequence stars on the SFR of the CMZ. A summary and outlook is given in Section~\ref{summary}. 

\section{Models}
\label{model}
To investigate whether main-sequence stars ({\bf Stage 2+}) embedded in an ambient density medium could mimic deeply embedded YSOs ({\bf Stage 0/1}) and match the measured brightness profile of the real observation in MIPS 24\,$\mu$m, we performed radiative transfer calculations. We set up 660 models for different spectral types and evolutionary stages in an ambient medium with different dust and density properties. We used the 3-d Monte Carlo radiative transfer code \textsc{Hyperion} \citep{2011:Robitaille} to compute the temperature distribution and create synthetic images. By further modeling the effects of convolution with arbitrary PSFs, transmission curves, finite pixel resolution, noise and reddening, our radiative transfer models are then directly comparable to real observations. Our synthetic pipeline \textsc{The FluxCompensator} will be made publicly available in the future\footnote{For more information about \textsc{Hyperion}, and to sign up to be notified once the \textsc{The FluxCompensator} package used here is available, visit \url{http://www.hyperion-rt.org}.}.

\subsection{Spectral types \& stages of evolution in an ambient density environment}
\label{spectral}

We modeled main sequence and young embedded O, B and A stars, with temperatures ranging from 44,500 to 8200\,K, using in both cases the stellar atmosphere models of \cite{2004:Castelli} as the central stars. We modeled the circumstellar density structure of the YSO models using a rotationally flattened envelope profile \citep{1976:Ulrich}, with gas infall rates from $3\times 10^{-4}\,\mbox{M}_{\odot}\,\mbox{yr}^{-1}$ to $3\times 10^{-8}\,\mbox{M}_{\odot}\,\mbox{yr}^{-1}$ determined from the scaling of the envelope density, an outer radius 1.5\,pc, and a centrifugal radius at 100\,AU. We assumed a gas-to-dust ratio of 100. The sublimation temperature, above which dust is removed, was set to 1600\,K. For all spectral types, we calculated 10 YSOs models and one additional model without an envelope (but with the constant density ambient medium), representing a main-sequence object in our simple approach. We use infall rate and stellar mass to classify every model as {\bf Stage 0/1} or {\bf Stage 2+} using the \cite{2006:Robitaille} formalism described in Section~\ref{intro}. The stellar data listed in Table~\ref{stellar_data} from Appendix E of \cite{1996:Carroll} was used to determine the stellar radii and luminosities.

\begin{table}[t]
	\caption{Stellar data used in radiative transfer setup.}
	\label{stellar_data}
		\begin{center}
		\leavevmode
		\begin{minipage}{\textwidth}
			\begin{tabular*}{0.48\textwidth}{cD{.}{.}{0}D{.}{.}{0}D{.}{.}{1}D{.}{.}{1}}
				\hline\hline
    				\multicolumn{1}{c}{\ \ \ SpT\ \ \ } 	& 	\multicolumn{1}{c}{\ \ \ T [K]\ \ \ }	&	\multicolumn{1}{c}{\ \ \ L [L$_\odot$]\ \ \ } 	&	\multicolumn{1}{c}{\ \ \ R [R$_\odot$]\ \ \ } 	& 	\multicolumn{1}{c}{\ \ \ M [M$_\odot$]}	\\
				\hline
				O5		&	44500	&	790000			&	15				&	60.			\\
				O6		&	41000	&	420000			&	13				&	37.			\\
				O8		&	35800	&	170000			&	11				&	23.			\\
				B0		&	30000	&	52000			&	8.4				&	17.5			\\
				B1		&	25400	&	16000			&	6.5				&	13.\footnotemark[1]			\\
				B2		&	22000	&	5700			&	5.2			    &	10.\footnotemark[1]			\\
				B3		&	18700	&	1900			&	4.2			    &	7.6			\\
				B5		&	15400	&	830				&	4.1				&	5.9			\\
				B8		&	11900	&	180				&	3.2				&	3.8			\\
				A0		&	9520	&	54				&   2.7             &   2.9\\
				A5		&	8200	&	14				&   1.9             &   2.0\\
 				\hline
  			\end{tabular*}
  			\footnotetext[1]{values from interpolation of the stellar data}
  			\end{minipage}
		\end{center}
\end{table}

We placed all models within a surrounding ambient medium with a constant density $\rho_0$. We used three different ambient density environments $\rho_{0}=[1,\ 3,\ 10]\times 10^{-21}~\mbox{g\,cm}^{-3}$, which are roughly within the number density range of $[10^3; 10^4]~\mbox{cm}^{-3}$ found for the CMZ (see \citealp{2011:Molinari}, \citealp{2013:Longmore}). For the ambient medium, we also assumed a gas to dust ratio of 100.

\begin{table*}[ht!]
	\caption{Information about telescopes and detectors.}
	\label{detectors}
		\begin{center}
		\leavevmode
		\begin{minipage}{5in}

			\begin{tabular}{lD{.}{.}{3}ccc}
				\hline\hline
				\multicolumn{1}{c}{name} 			& 
    				\multicolumn{1}{c}{zero-point} 		&
    				\multicolumn{1}{c}{filter}			&
    				\multicolumn{1}{c}{pixel size} 		&	    				\multicolumn{1}{c}{PSF}				\\
    				
    				\multicolumn{1}{c}{\ } 			&	
    				\multicolumn{1}{c}{[Jy]} 		&
    				\multicolumn{1}{c}{\ } 			&
    				\multicolumn{1}{c}{[arcsec]} 	&
    				\multicolumn{1}{c}{\ } 		\\
				\hline
UKIDSS K	
&631.\footnotemark[5]		
&\cite{2006:Hewett}
&0.4\footnotemark[1]
&Gaussian
\\

IRAC 8$\mu$m	
&64.9\footnotemark[2]		
&\cite{2004:Quijada}
&1.2\footnotemark[2]
&\cite{2011:Aniano}\footnotemark[8]
\\

MIPS 24 $\mu$m
&7.17\footnotemark[3]			
&MIPS Handbook\footnotemark[3]
&2.4\footnotemark[3]
&Empirical
\\

PACS 70 $\mu$m			
&0.78\footnotemark[6]
&Herschel Science Center\footnotemark[7]
&3.2\footnotemark[4]
&\cite{2011:Aniano}\footnotemark[8]
\\
\hline
\end{tabular}


\footnotetext[1]{UKIDSS Handbook: \url{http://ukidss.org/technical/technical.html}}
\footnotetext[2]{IRAC Handbook: \url{http://irsa.ipac.caltech.edu/data/SPITZER/docs/irac}}
\footnotetext[3]{MIPS Handbook: \url{http://irsa.ipac.caltech.edu/data/SPITZER/docs/mips/}}
\footnotetext[4]{PACS Handbook: \url{http://herschel.esac.esa.int/Docs/PACS/html/
pacs_om.html}}
\footnotetext[5]{\cite{2006:Hewett}}
\footnotetext[6]{\url{http://svo2.cab.inta-csic.es/theory/fps/index.php?mode=browse&gname=Herschel}}
\footnotetext[7]{\url{https://nhscsci.ipac.caltech.edu/sc/index.php/Pacs/FilterCurves}}
\footnotetext[8]{\url{http://www.astro.princeton.edu/~ganiano/Kernels.html}}
\end{minipage}
\end{center}
\end{table*}

\subsection{Dust properties}
\label{dust}
For every combination of parameters, described in Section~\ref{spectral}, we run the model for two different sets of dust properties. The first was the Milky-Way dust from \cite{2003:Draine} with R$_V=5.5$ and b$_c=30\,\mbox{ppm}$, where R$_V$ is the ratio of the visual extinction to reddening magnitude, and b$_c$ is the concentration of carbon atoms in the medium. \cite{2001:Weingartner} and \cite{2003:Draine} favour this combination for the Milky Way and point out that it best reproduces the conditions in the galactic center observed by \cite{1996:Lutz}. In the second configuration, in order to test the effect of poly-aromatic hydrocarbons (PAH), we additionally used the dust properties by \cite{2007:Draine} and as used in \cite{2012:Robitaille} with a mixture of 5.9\,\% ultra-small grains, 13.5\,\% very small grains and 80.6\,\% big grains with R$_V=3.1$ and b$_c=52\,\mbox{ppm}$. 

\subsection{Realistic synthetic observations}
\label{synth-obs}
The synthetic images and spectral energy distributions (SEDs) computed by \textsc{Hyperion} are not directly comparable with photometric observations. In order to make these radiative transfer ``observations'' as realistic as possible, we developed a synthetic observations pipeline called \textsc{The FluxCompensator}. In what follows we will describe it briefly. 

For every model we produced realistic synthetic images as they were observed in four bands: the $K$ filter of the UKIDSS Galactic plane survey \citep{2008:Lucas}, and the IRAC 8\,$\mu$m, MIPS 24\,$\mu$m and PACS 70\,$\mu$m bands described in Section~\ref{intro}. Further information on these filters is provided in Table~\ref{detectors}, and in the UKIDSS, \emph{Spitzer} and \emph{Herschel} documentation. We convolved the synthetic images from \textsc{Hyperion} with the respective PSF after adjusting the pixel resolution. The original PSF files from \cite{2011:Aniano} were used, except the MIPS 24\,$\mu$m PSF, which was directly derived from the observations.

We applied a filter convolution with the corresponding transmission functions from the detectors, and we accounted for reddening using the extinction law provided by \cite{1994:Kim} and an optical extinction value of A$_V=20$\,mag. Estimates for the visual extinction towards the central molecular zone typically range from 20 to 30\,mag (see \citealp{2011:Geballe}), although the higher values likely include a contribution from extinction local to Sgr A; therefore, we assumed a value of 20\,mag. Similarly to \cite{2013:Longmore} and \cite{2009:Yusef} we assume a distance of 8.5\,kpc to the CMZ. After generating realistic synthetic images for every model, we additionally measured the magnitudes and peak surface brightnesses\footnote{These and further parameters are provided in the Appendix and accessible in the online material.}. We calculated the total flux from the synthetic images with a field of view of 50.4\arcsec\ x 50.4\arcsec\ for UKIDSS~$K$, IRAC 8\,$\mu$m and MIPS 24\,$\mu$m. In all cases, the flux derived in this way is equivalent within  1\,\% to the total integrated flux of the sources, that would be measured by standard photometry techniques such as PSF-fitting or aperture photometry (with aperture correction). We then converted these fluxes to magnitudes for the purpose of comparing these to observations. For the peak surface brightness we interpolate with 2D cubic spline interpolation to extract the value at the real peak.

With the \textsc{FluxCompensator}, it is also possible to add noise. However, in order to account for a similar background as the one present in the real observations, we did not add synthetic noise. Instead it is possible to add the realistic synthetic image to the real observations (see Figure~\ref{add2obs}), so that the models are directly comparable with real objects. This comparison is meaningful because for the ambient volume density we chose values comparable to measured average densities in the CMZ (see Section~\ref{spectral}).

\newpage
\section{Results}
\label{results}

\subsection{Real observations vs. realistic synthetic observations}
\label{compare obs}

\begin{figure}
\includegraphics[width=0.5\textwidth]{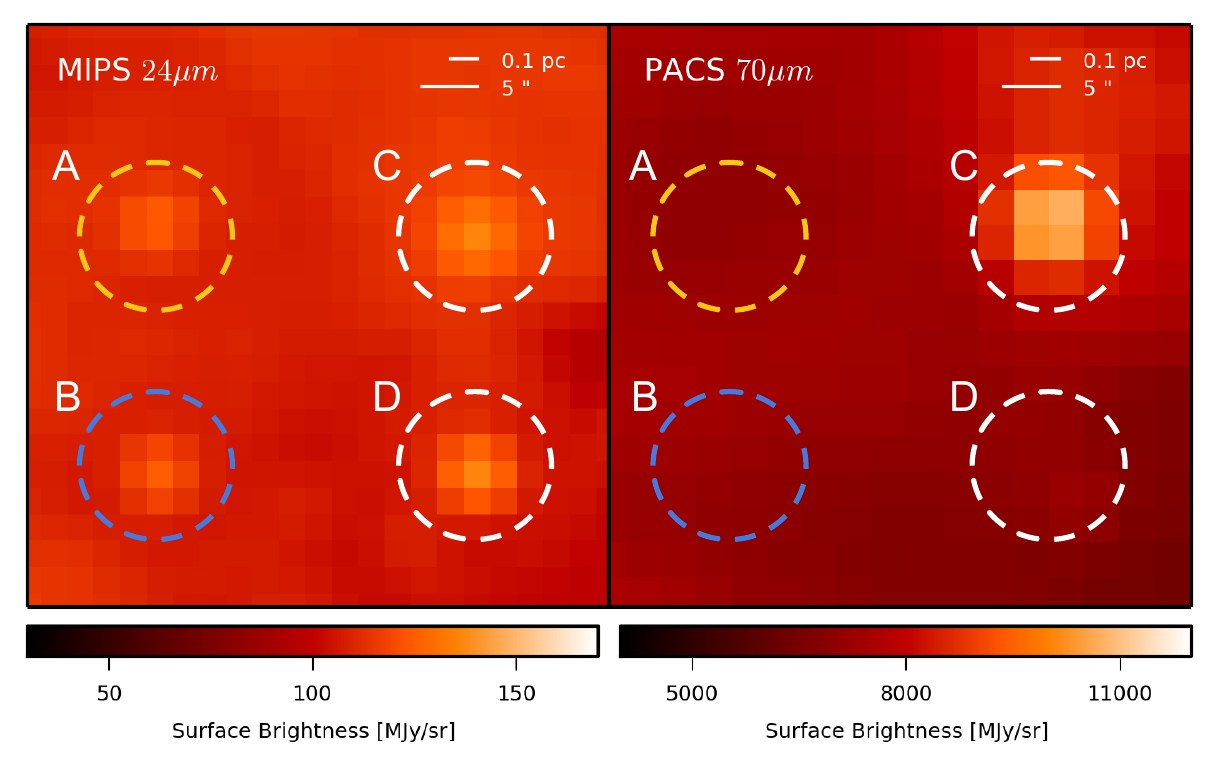}
\caption{\label{add2obs}A patch of the CMZ observed by MIPS 24\,$\mu$m (left) and PACS 70\,$\mu$m (right). The real object (A) is marked with a yellow dashed circle, the synthetic {\bf Stage 2+} source (B) by a blue circle and the two synthetic {\bf Stage 0/1} sources (C, D) by white circles. In MIPS 24\,$\mu$m the synthetic observations shown agree with the real observation. The synthetic {\bf Stage 2+} source as well as the lower synthetic {\bf Stage 0/1} source agree with the PACS 70\,$\mu$m observation of the real source.}
\end{figure}

In this section, we compare three model objects, as described in Section~\ref{model}, added to observations to compare to a real source classified as a YSO. In Figure~\ref{add2obs} we placed three synthetic observations (A0 {\bf Stage 2+}, B5 and B8 {\bf Stage 0/1}) next to a classified YSO of {\bf Stage 0/1} by \cite{2009:Yusef} at equatorial coordinates $\alpha_{\rm J2000} = 17^h44^m26.835^s$, $\delta_{\rm J2000} = -29^{\circ}15^{\prime}21.05^{\prime\prime}$ (yellow circle), which is clearly visible in MIPS 24\,$\mu$m, but with no counterpart in the \emph{Herschel} observations.

\begin{figure}
\centering
\includegraphics[width=0.5\textwidth]{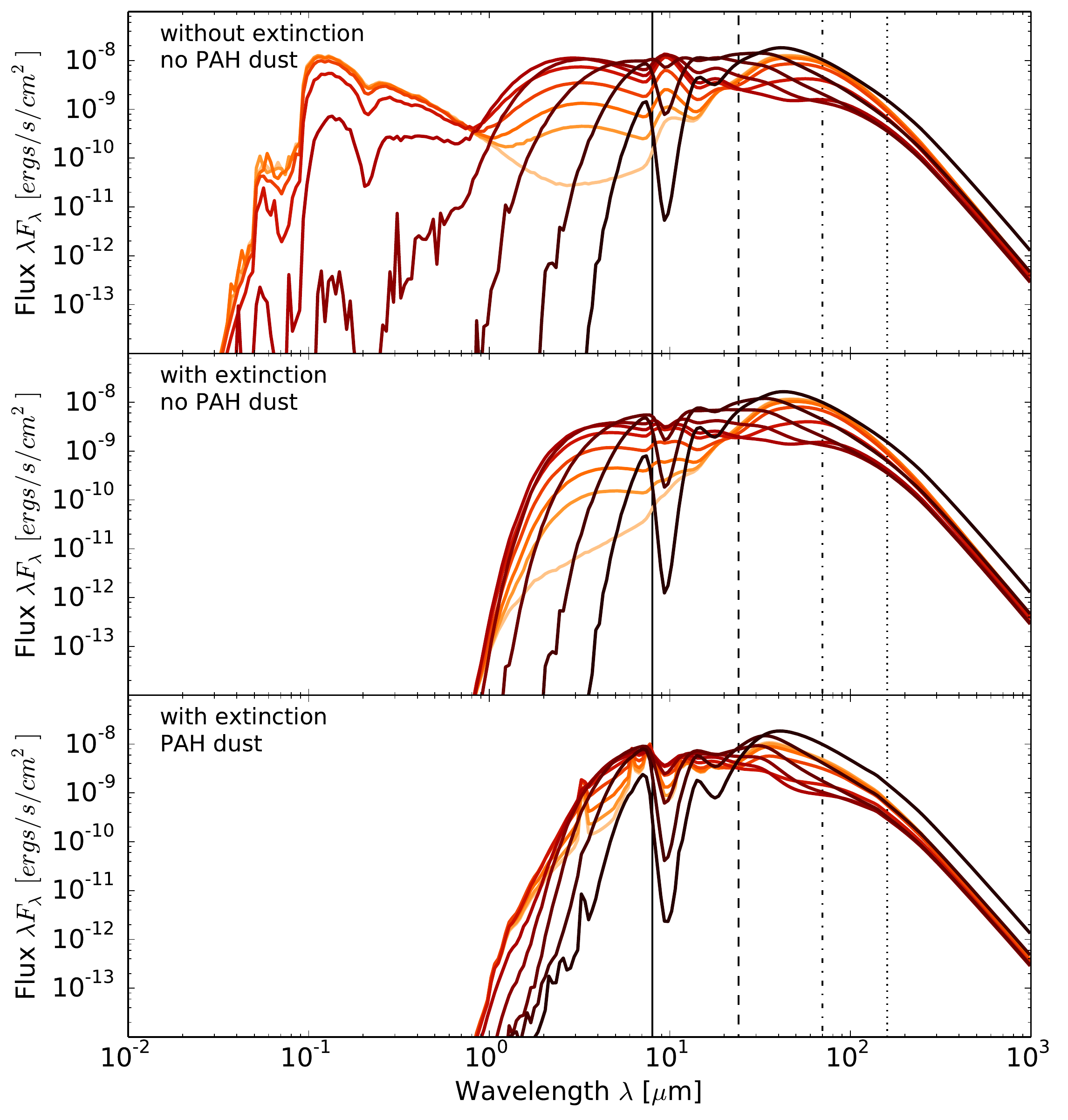}
\caption{\label{SED_pah}Evolution of the synthetic SEDs at different evolutionary stages of a B0 star embedded within an ambient medium of density $\rho_{0}=10^{-21}\,\mbox{g\,cm}^{-3}$ considering the effects of extinction and PAH dust. Vertical lines (solid: IRAC 8\,$\mu$m, dashed: MIPS 24\,$\mu$m, dot-dashed: PACS 70\,$\mu$m, dotted: PACS 160\,$\mu$m), SEDs from black to yellow ($3\times 10^{-4}\,\mbox{M}_{\odot}\,\mbox{yr}^{-1}$, $10^{-4}\,\mbox{M}_{\odot}\,\mbox{yr}^{-1}$, $3\times  10^{-5}\,\mbox{M}_{\odot}\,\mbox{yr}^{-1}$, $10^{-5}\,\mbox{M}_{\odot}\,\mbox{yr}^{-1}$, $3\times 10^{-6}\,\mbox{M}_{\odot}\,\mbox{yr}^{-1}$, $10^{-6}\,\mbox{M}_{\odot}\,\mbox{yr}^{-1}$, $3\times 10^{-7}\,\mbox{M}_{\odot}\,\mbox{yr}^{-1}$, $10^{-7}\,\mbox{M}_{\odot}\,\mbox{yr}^{-1}$, $3\times 10^{-8}\,\mbox{M}_{\odot}\,\mbox{yr}^{-1}$, main-sequence star with no envelope).}
\end{figure}

\begin{figure}[ht]
\vspace*{-0.1cm}
\centering
\includegraphics[width=0.5\textwidth]{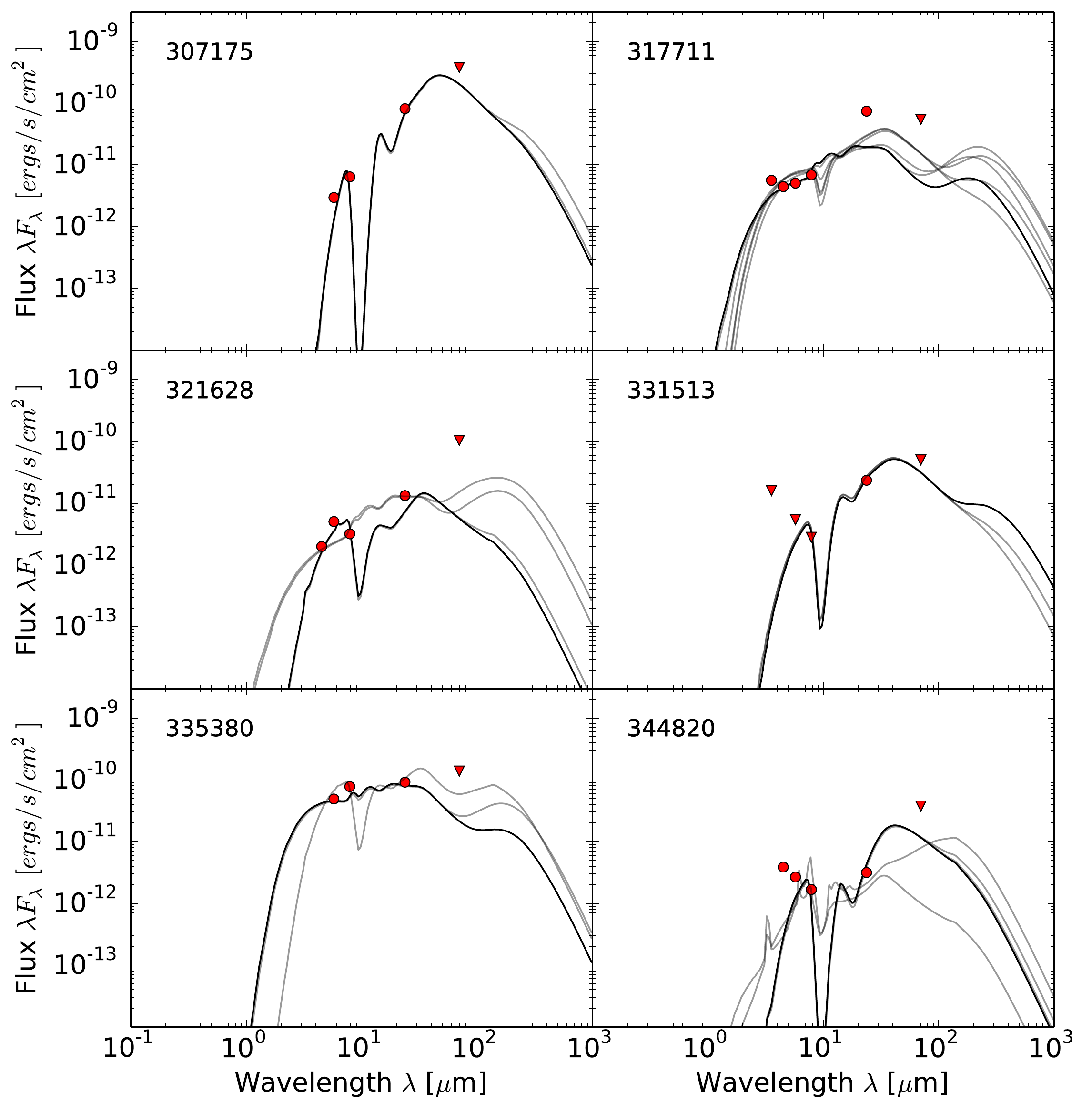}
\caption{\label{SED_obs}Real SEDs (circles) of six classified YSOs shown in Figure~\ref{Milky Way}. The black and gray solid lines represent the synthetic SEDs from the modeling with the best and acceptable $\chi^2$ fits, respectively. The triangles represent 5$\sigma$ upper limits in PACS 70\,$\mu$m (and for object 331513 also in the IRAC bands).}
\end{figure}

Our radiative transfer model of an embedded B5 YSO of {\bf Stage 0/1} with $\dot{M}=3\times 10^{-4}\,\mbox{M}_{\odot}\,\mbox{yr}^{-1}$ (C: upper white circle) matches the real object in MIPS 24\,$\mu$m (A: yellow circle), but also produces a source detected in PACS 70\,$\mu$m. On the other hand, the B8 {\bf Stage 0/1} model with $\dot{M}=10^{-4}\,\mbox{M}_{\odot}\,\mbox{yr}^{-1}$ (D: lower white circle) has only a hardly detectable counterpart in PACS 70\,$\mu$m, while matching the MIPS observation. Our model of a more evolved source, an A0 {\bf Stage 2+} with $\dot{M}=3\times 10^{-7}\,\mbox{M}_{\odot}\,\mbox{yr}^{-1}$ has no counterpart in PACS 70\,$\mu$m (B: blue circle). Therefore, the observed source can be explained by both {\bf Stage 0/1} and {\bf Stage 2+} models, and so may not be as young as 1\,Myr.

\subsection{Evolution in the Spectral Energy Distribution}
\label{SED}
In Figure~\ref{SED_pah}, we show the evolution in the SED of a B0 star from deeply embedded to main sequence object, for three dust configurations: with regular dust and no extinction; with regular dust and extinction; and with PAH dust and extinction. Naturally, the extinction affects more strongly the near-infrared (NIR) bands. The PAHs add emission features in the mid-infrared (MIR), but does not change too much above 24\,$\mu$m. The far-infrared (FIR) remains almost unaffected by both extinction and PAH dust emission. 

\begin{figure*}
\centering
\includegraphics[width=\textwidth]{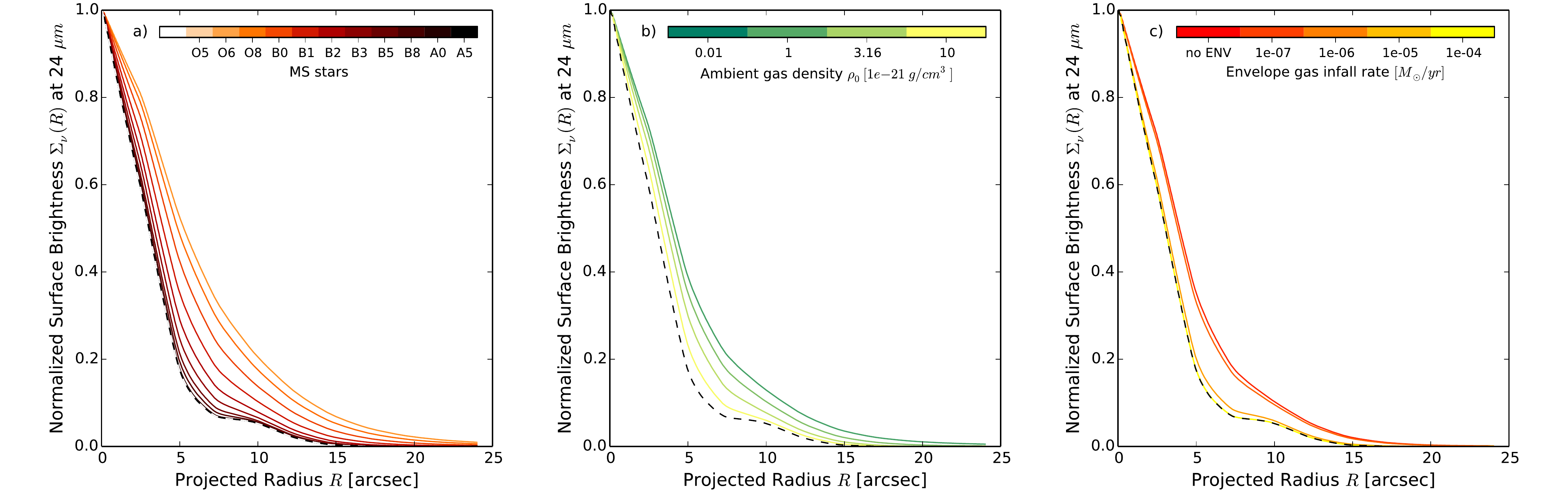}
\caption{\label{Sigma}MIPS 24\,$\mu$m radial brightness profiles after convolution with the PSF (shown as the dashed line). a) main-sequence objects without cirumstellar material, ambient medium gas density $\rho_{0}=1\times 10^{-21}\,\mbox{g\,cm}^{-3}$, no PAH dust b) B0 main-sequence star without cirumstellar material but with varying ambient medium $\rho_{0}$ c) B0 star with various infall rates embedded in an ambient medium with gas density $\rho_{0}=1\times 10^{-21}\,\mbox{g\,cm}^{-3}$, no PAH dust. Models with hotter central stars and less dense circumstellar or ambient medium are more resolved that models with cooler central objects and/or models with denser circumstellar and/or ambient medium.}
\end{figure*}

Below we describe the evolution of the SED as a function of evolutionary stage. The evolution of the SED from a main sequence source to a deeply embedded objects is best observed for regular dust and extinction (middle panel of Figure~\ref{SED_pah}). For a main-sequence object (with no envelope, yellow line), which is surrounded by an ambient medium, most of the mass is located at larger radii where the temperature is cold, and the mass of hotter material is low. This explains the lack of MIR emission. For the models with increasing infall rate, more material is added in the hotter regions closer to the star. As long as the dust is optically thin, the temperature of the dust in the inner regions stays constant, but the mass of the dust at these temperatures increases. The flux from this heated dust in the center is emitted in the NIR and MIR, which causes the rise at these wavelengths (e.\,g.~infall rate $10^{-6}\,\mbox{M}_{\odot}\,\mbox{yr}^{-1}$). NIR photons escape the system since the probability of re-absorption is too small. A higher accretion rate increases the envelope density, and the envelope starts to become optically thick to the stellar radiation at optical wavelengths. The temperature in the outer regions drops, resulting in a drop in the FIR emission. The emission in NIR and MIR goes down, once the envelope is also no longer optically thin at that wavelength (see infall rate $>3\times 10^{-6}\,\mbox{M}_{\odot}\,\mbox{yr}^{-1}$), and the FIR emission rises again due to the absorbed radiation getting re-emitted. 

For the six classified YSOs in the panels of Figure~\ref{Milky Way} we now plot the measured SEDs in Figure~\ref{SED_obs}. We used the available IRAC fluxes published in the point-source catalog of \cite{2008:Ramirez}, and extracted the MIPS 24\,$\mu$m from the point-source catalog provided by \cite{2009:Hinz}. These measured fluxes were the same as used by \cite{2009:Yusef}. We computed the 5$\sigma$ upper limits for PACS 70\,$\mu$m  (triangles) after removing the background. We assumed an average error of 10\,\% in the fluxes as used by \cite{2007:Robitaille}. The black solid line shows the best fit $\chi_{best}^2$ with the lowest $\chi^2$. The grey solid lines represent the synthetic model SEDs for which $(\chi^2 - \chi_{best}^2)/N_{sample}<3$ as defined in \cite{2007:Robitaille}. Object 307175 is very well fitted by a B5 {\bf Stage 0/1} models. For object 331513, we found from inspecting the images that the IRAC 8\,$\mu$m source may not be related to the MIPS 24\,$\mu$m source since it is not well aligned, so we use the IRAC fluxes as 5$\sigma$ upper limits (triangles). Object 344820 is not well fit. It has a very weak MIPS 24\,$\mu$m source. We remeasured its total flux and found 21.5 instead of 26.7\,mJy, which improves the fit marginally, but since the IRAC fluxes are still not well fit, and are typical of an unextincted source, they may originate from an unrelated foreground object. Objects 317711, 321628, 335380 and 344820 can be either fitted by {\bf Stage 0/1} or {\bf Stage 2+} objects of spectral type B5, B8, A0, A5 and B8, B5, A0, respectively. Hence, that SED fitting alone can not be used to distinguish between true YSOs and more evolved objects.

\subsection{Evolution in the radial profiles}
\label{radprofile}
By inspecting the YSOs classified by \citet{2009:Yusef} in the 24\,$\mu$m images, we found that some objects appear to be resolved, i.\,e.~some sources are 
slightly larger than point sources. Therefore, we explore the radial brightness profiles of our synthetic images in MIPS 24\,$\mu$m. 

In Figure~\ref{Sigma}, a compilation of normalized profiles of the realistic synthetic observations are shown. The dashed line represents the PSF of a perfect point source. Its full-width-half-maximum (FWHM) is roughly 6\,\arcsec. Without an envelope (Figure~\ref{Sigma}\textcolor{blue}{a}) the sources become more extended for higher temperature stars. The low density of the surrounding medium enhances this effect (Figure~\ref{Sigma}\textcolor{blue}{b}), while the presence of circumstellar dust reduces the width of the profile (Figure~\ref{Sigma}\textcolor{blue}{c}). Resolved sources could therefore be earlier type stars and/or stars embedded in a low density environment (either circumstellar and/or ambient medium). 

\subsection{Distinguishing embedded YSOs from more evolved objects}
\label{dia}

In this Section, based on the observational properties of our models such as MIPS 24\,$\mu$m magnitude, MIPS 24\,$\mu$m angular size and MIR color, we explore how to distinguish between true YSOs ({\bf Stage 0/1}) from more evolved objects ({\bf Stage 2+}) in the CMZ.
\begin{figure}
\includegraphics[trim={0.95cm 0cm 1.5cm 1.cm}, clip ,width=0.5\textwidth]{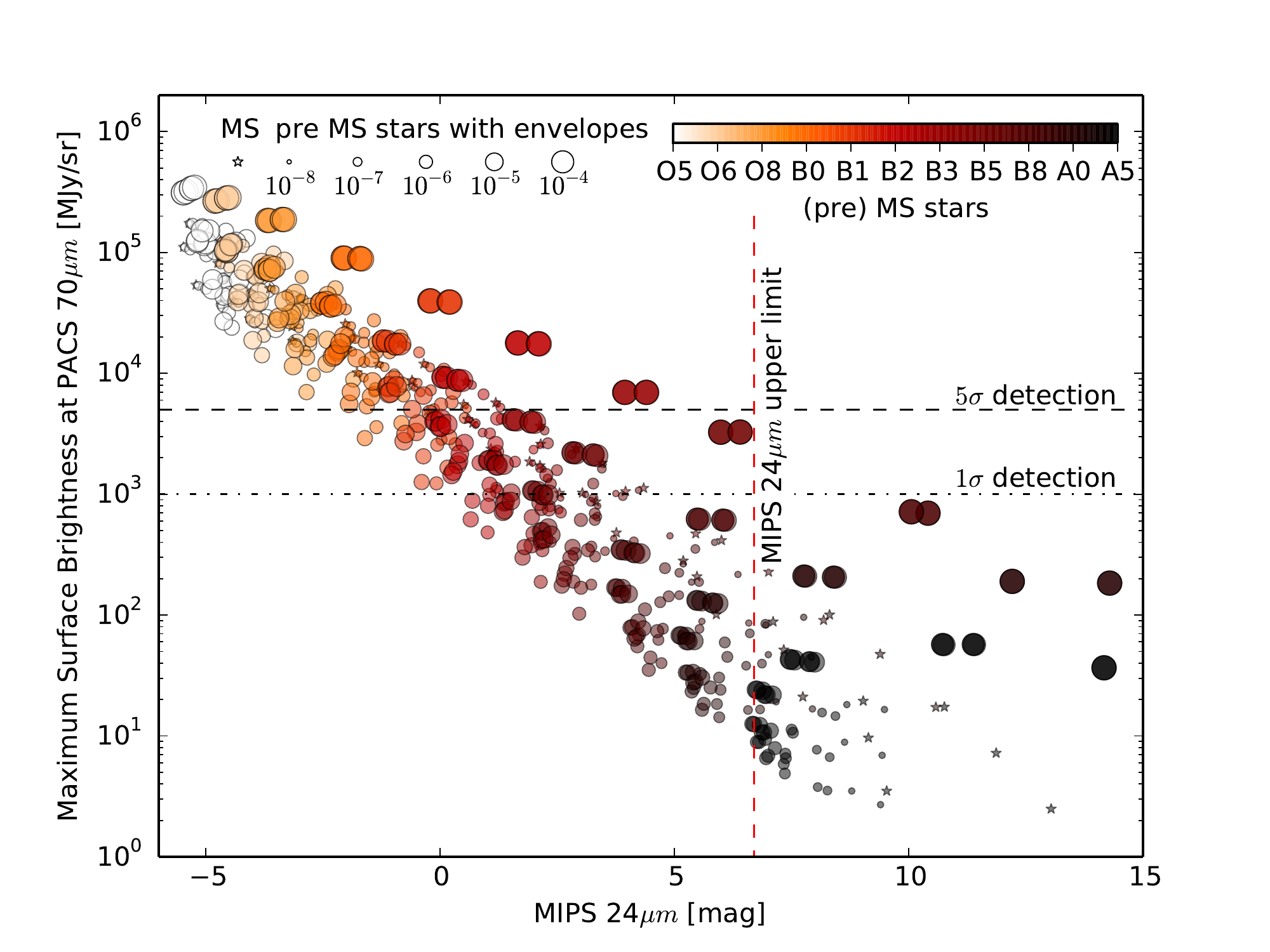}
\caption{\label{fluxSpT}PACS 70\,$\mu$m peak surface brightness and MIPS 24\,$\mu$m magnitudes for all the models, color coded by spectral type.
The detection limits of HiGal and MIPSGal in the CMZ are shown as horizontal and vertical lines.}
\end{figure}
\begin{figure*}
\includegraphics[trim={0.95cm 0cm 1.5cm 1.5cm}, clip ,width=0.5\textwidth]{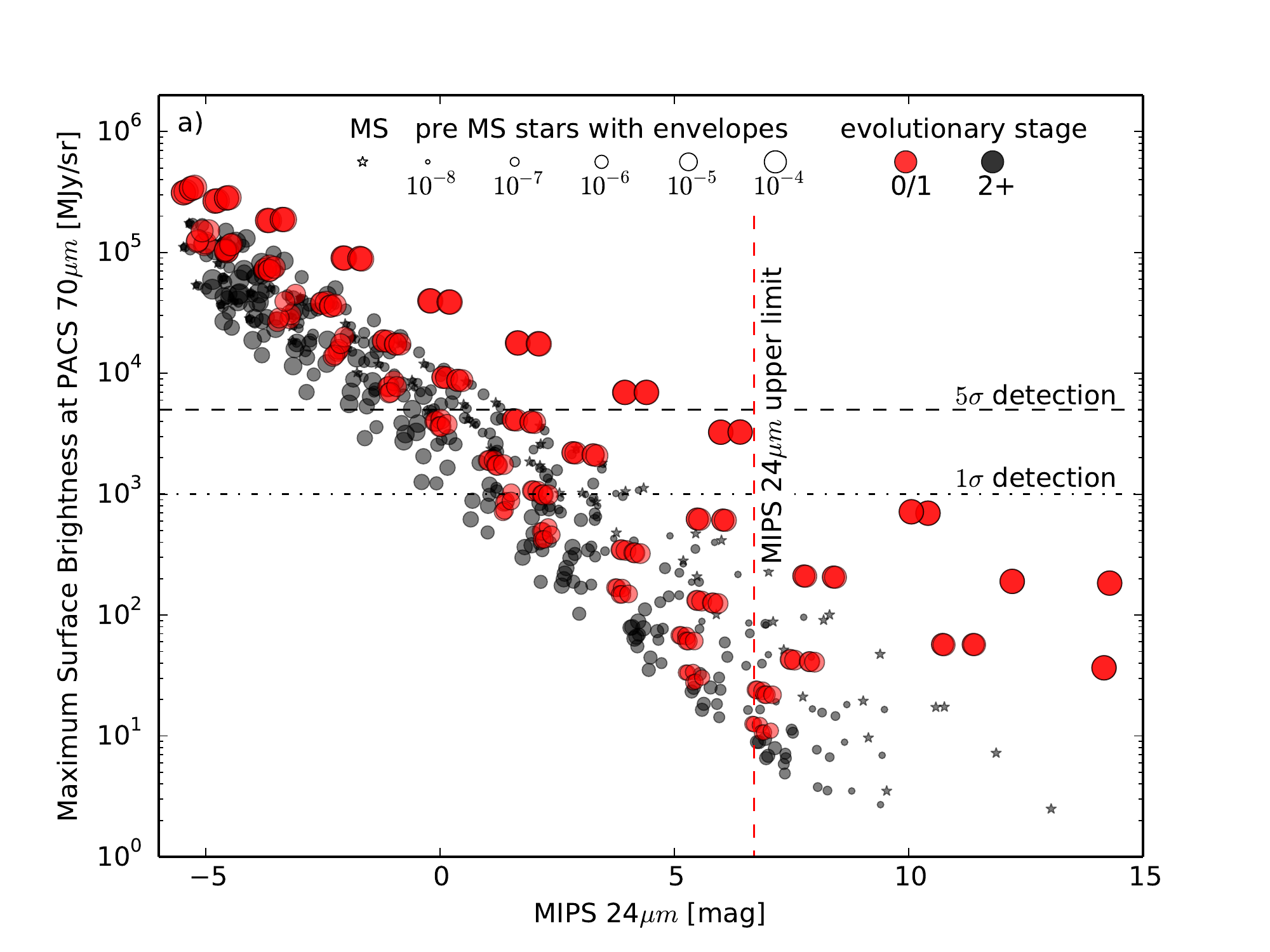}
\includegraphics[trim={0.95cm 0cm 1.5cm 1.5cm}, clip ,width=0.5\textwidth]{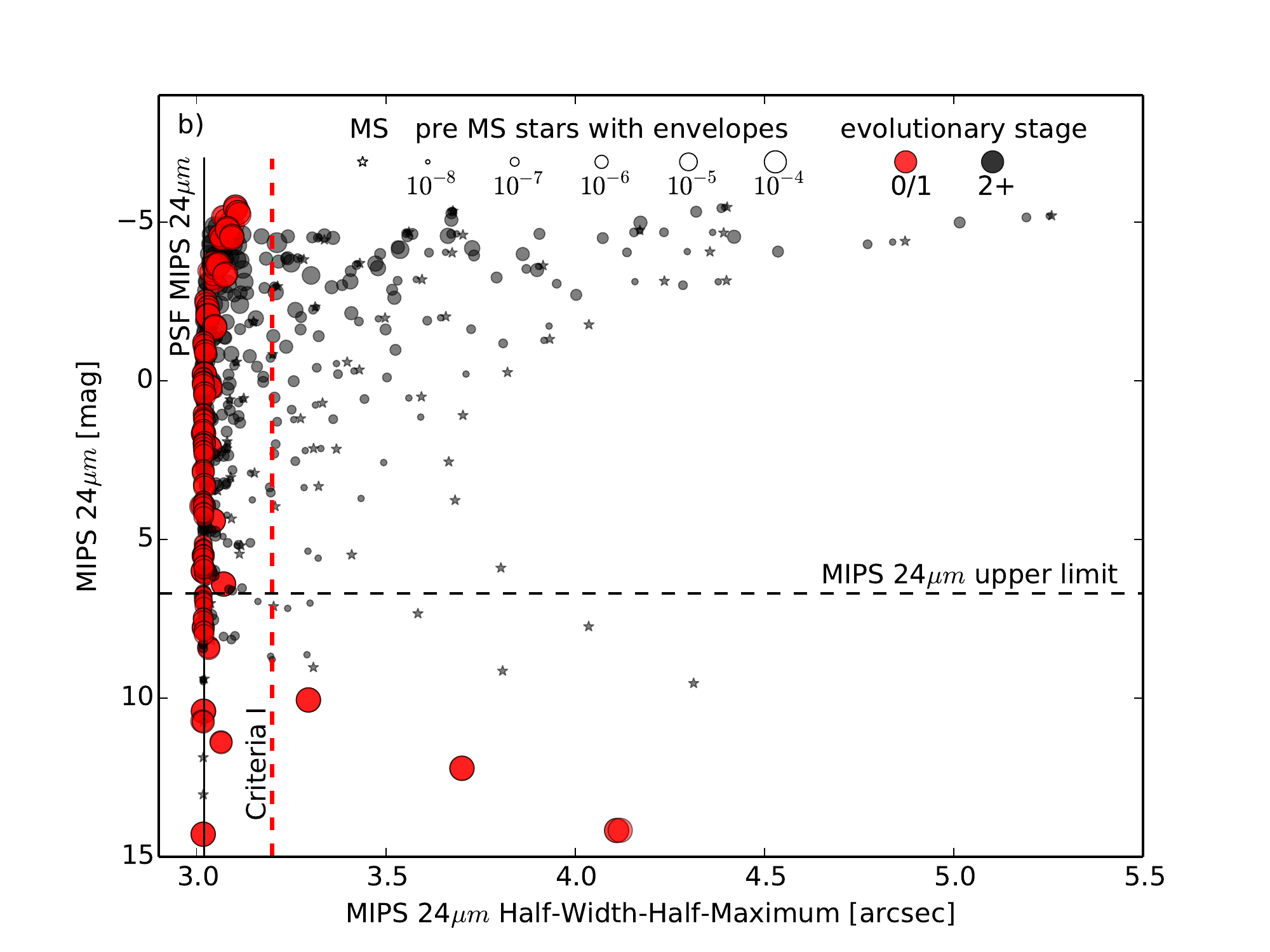}
\includegraphics[trim={0.95cm 0cm 1.5cm 1cm}, clip ,width=0.5\textwidth]{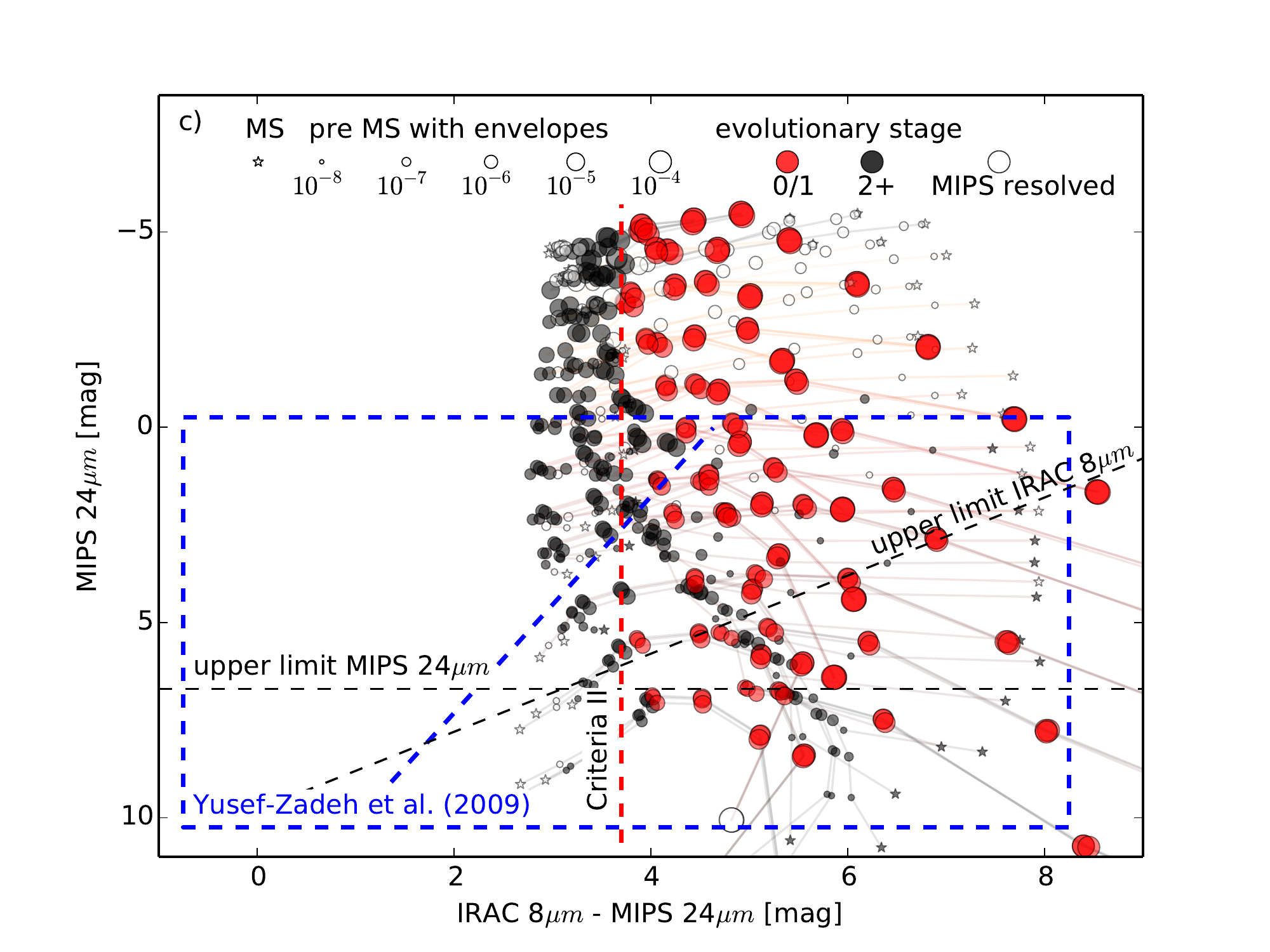}
\includegraphics[trim={0.95cm 0cm 1.5cm 1cm}, clip ,width=0.5\textwidth]{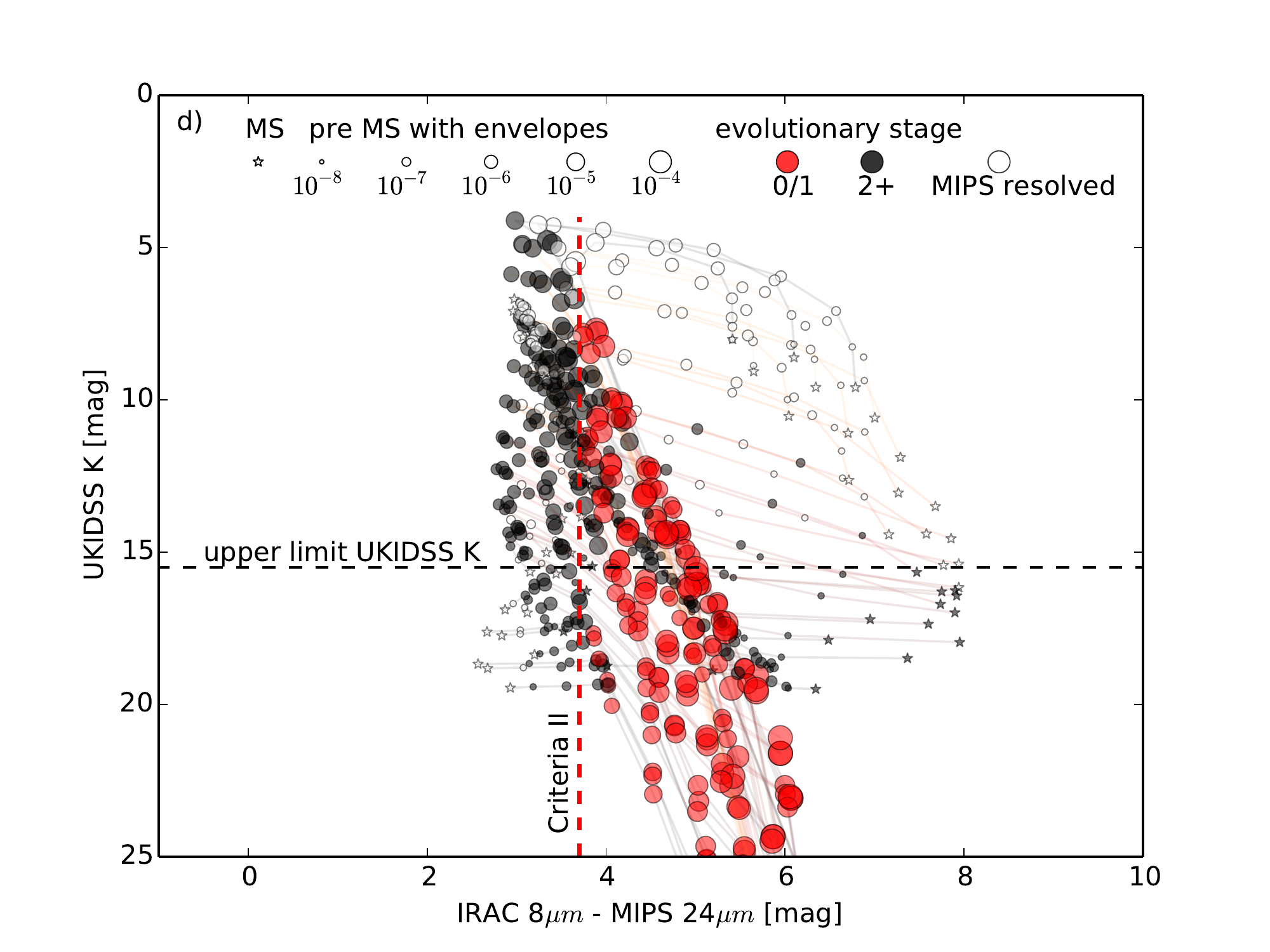}
\caption{\label{flux}Diagnostic diagrams extracted from our synthetic observations: 
a) Peak PACS 70\,$\mu$m surface brightness vs. and MIPS 24\,$\mu$m magnitude diagram. The red line shows the MIPS 24\,$\mu$m and the black lines show the PACS 70\,$\mu$m detection limits. 
b) MIPS 24\,$\mu$m HWHM vs. MIPS 24\,$\mu$m. Red dashed line shows our criteria to distinguish between resolved and unresolved objects, black dashed line is the MIPS 24\,$\mu$m upper limit. 
c) Color-magnitude diagram in the MIR. The blue dashed box represents the plotting limits of \cite{2009:Yusef}, blue dashed diagonal line is the empirical criteria by \cite{2009:Yusef} to distinguish AGB stars from YSOs. Black dashed lines are upper limits in MIPS 24\,$\mu$m and IRAC 8\,$\mu$m. The red vertical dashed line separates more evolved objects from the sample.  
d) Color-magnitude diagram MIR vs. NIR. The black dashed line is upper limit in UKIDSS K. The red vertical dashed line separates more evolved objects from the sample. (red: {\bf Stage 0/1}, black: {\bf Stage 2+}, white: MIPS resolved)}
\end{figure*}
\subsubsection{Detection in MIPS 24\,$\mu$m and PACS 70\,$\mu$m}
Using the source counts as a function of magnitude of the point source catalogues, we determined approximate detection limits for the western part of the CMZ, where most of the YSOs classified by \cite{2009:Yusef} are located. For UKIDSS~$K$, IRAC 8\,$\mu$m and MIPS 24\,$\mu$m we found, respectively, upper limits of 14.8\,mag, 8.9\,mag and 6.0\,mag for a completeness of 90\,\%, and 15.5 \,mag, 9.8\,mag and 6.7\,mag for a completeness of 50\,\%, within an error of about 0.25\,mag, which is the bin width used to construct the magnitude histograms. For PACS 70\,$\mu$m we estimated a $5\sigma$ surface brightness detection to be at roughly 5000\,MJy/sr.
In Figure~\ref{fluxSpT}, we show the PACS 70\,$\mu$m peak surface brightness versus the MIPS 24\,$\mu$m magnitude for the 660 modelled objects, indicating the different spectral types and envelope infall rates. A total of 567 model objects can be detected in MIPS 24\,$\mu$m as indicated by the vertical dashed line in Figure~\ref{fluxSpT}. The MIPS 24\,$\mu$m detection limit corresponds to a completeness limit of B5 in spectral type. On the other hand, the PACS 70\,$\mu$m $5\sigma$ detection limit (also shown in Figure~\ref{fluxSpT}) translates into a completeness limit of O8 in spectral type, but also some evolutionary stages of B0 to B3 objects can be $5\sigma$-detected in PACS 70\,$\mu$m, while later types show no $5\sigma$-detected counterparts. 

In Figure~\ref{flux}\textcolor{blue}{a} the models from Figure~\ref{fluxSpT} are classified with the Stage formalism described in Section~\ref{intro}. As we can see, {\bf Stage 0/1} models lie above and below the $5\sigma$-detection limit in PACS 70\,$\mu$m. Therefore, detection or non-detection of sources at PACS 70\,$\mu$m does not allow us distinguish true YSOs from more evolved objects in the CMZ, disproving the hypothesis put forward in Section~\ref{intro} that objects only seen at MIPS 24\,$\mu$m and not detected at PACS 70\,$\mu$m would not be YSOs.

\subsubsection{Half-width-half-maximum in MIPS 24\,$\mu$m}
\label{hwsize}
To explore the effects in the radial profiles described in Section~\ref{radprofile}, we need the half-width-half-maximum (HWHM) of all models in MIPS 24\,$\mu$m. We developed a tool which can fit PSF models for extended sources, which consist of Gaussian profiles convolved with the MIPS 24\,$\mu$m PSF. The fit can be carried out manually in order to ensure an optimal fit in regions of complex background. The profile with the best fit is used to calculate the `observed' HWHM, which is roughly $\sqrt{\mbox{HWHM}_{PSF}^2 + \mbox{HWHM}_{Gauss}^2}$. We fit the synthetic sources with 100 profiles with combined HWHMs ranging from 3 to 6\arcsec\ (1 HWHM to 2 HWHM). With this PSF fitting method it was also possible to extract total integrated fluxes. In Section~\ref{SFRcorr} we use this technique on real observations.

In Figure~\ref{flux}\textcolor{blue}{b}, we plot the MIPS 24\,$\mu$m magnitude vs. the MIPS 24\,$\mu$m HWHM. About half of the objects lie exactly on the PSF (HWHM = 3.02\,\arcsec) and are therefore unresolved. We find that most resolved objects are have an inverse timescale $\dot{\mbox{M}}_{gas}/\mbox{M}_\star$ less than $10^{-6}\,\mbox{yr}^{-1}$ and are therefore not likely to be as young as assumed by e.\,g.~\cite{2009:Yusef}. 

We adopt the Stage formalism to explore whether the objects in the vicinity of the PSF can be disentangled. We found that 228 of the 660 model objects are in fact still in the envelope dominated phase ({\bf Stage 0/1}). We define a HWHM threshold of 3.2\,\arcsec\ to separate \emph{resolved} objects (HWHM $\geq 3.2$\arcsec) from \emph{unresolved} objects (HWHM $< 3.2$\arcsec). There are only nine resolved {\bf Stage 0/1} model objects, and these are all below the MIPS 24μm detection limit. Therefore, most of the resolved objects (165 of 174) are more evolved {\bf Stage 2+}. A total of 486 objects are unresolved and include objects from both Stages of evolution (219 {\bf Stage 0/1}, 267 {\bf Stage 2+}). Note that while seeing an extended source at 24\,$\mu$m likely indicates that a source is not truly young, unresolved sources can still be ambiguous. 

\subsubsection{MIR and NIR color-magnitude diagrams}
In what follows, we now have a closer look at the 481 unresolved model objects using color-magnitude diagrams, in order to understand to what extend we can distinguish between embedded YSOs from more evolved objects. Figure~\ref{flux}\textcolor{blue}{c} presents the IRAC 8\,$\mu$m - MIPS 24\,$\mu$m vs. MIPS 24\,$\mu$m color-magnitude diagram. \cite{2009:Yusef} used an empirical criterion (see blue dashed diagonal line in Figure~\ref{flux}\textcolor{blue}{c}) to distinguish AGB field stars from YSOs. Our modelling shows that this criterion alone (without accounting for resolved and unresolved objects) is not effective to separate true YSOs from more evolved objects, since these overlap in color-magnitude space. The evolutionary tracks in the color-magnitude diagram can be explained similarly to the SED evolution described in Section~\ref{SED}. Although there is some overlap, sources with IRAC 8\,$\mu$m - MIPS 24\,$\mu$m $<$ 3.7\,mag are always more evolved.

One can see that the MIR color-magnitude diagram alone (without accounting for resolved and unresolved objects) is not capable to completely distinguish between more evolved and deeply embedded YSOs. Combining bands in the NIR and the MIR seems more promising if the detection in $K$ is sensitive enough, as can be noted in Figure~\ref{flux}\textcolor{blue}{d}, which shows the IRAC 8\,$\mu$m - MIPS 24\,$\mu$m vs. UKIDSS~$K$ color-magnitude diagram. The $K$ band is dominated by stellar emission that is simply extincted, so it directly probes the envelope column density. Since the column density is very different between the main-sequence stars and ambient medium compared to with envelopes, all envelope dominated objects fall in one distinctive region in the diagram. 

\subsubsection{Criteria to select true YSOs}
\label{cond}

In summary, we have shown that the following criteria can be used to preferentially select truly young {\bf Stage 0/1} YSOs, based on the MIPS 24\,$\mu$m size and IRAC 8\,$\mu$m - MIPS 24\,$\mu$m color:
\begin{eqnarray}
\mbox{I} && \mbox{MIPS 24\,$\mu$m HWHM} < 3.2\,\mbox{\arcsec}\nonumber\\
\mbox{II} && \mbox{IRAC 8\,$\mu$m - MIPS 24\,$\mu$m} > 3.7\,\mbox{mag}\nonumber
\end{eqnarray}

Of the 660 model objects, 343 match these criteria. Of these, 219 are true YSOs models and instead of however 124  are models of more evolved objects. Nine objects are misclassified as more evolved but are true YSOs and lie below the MIPS 24\,$\mu$m detection limit. The remaining 308 are all models of more evolved objects. In summary, using these criteria on our models results in:
\begin{eqnarray}
\mbox{33.2\,\%} && \mbox{correctly classified {\bf Stage 0/1} objects}\nonumber\\
\mbox{46.7\,\%} && \mbox{correctly classified {\bf Stage 2+} objects}\nonumber\\
\mbox{18.8\,\%} && \mbox{misclassified as {\bf Stage 0/1}} \nonumber\\
\mbox{1.4\,\%} && \mbox{misclassified as {\bf Stage 2+}}\nonumber
\end{eqnarray}

When including protoplanetary disks ($r_{max}=1000\,\mbox{AU}$) to our YSO circumstellar geometry setup the above criteria are mostly unchanged. Overall models with discs are slightly less extended due to the increased mass of circumstellar material in the center. Therefore, the selection criteria I in MIPS 24\,$\mu$m is not quite as successful as for the runs without disks. For models with a disk, the classification fractions are as follows:
 \begin{eqnarray}
\mbox{33.2\,\%} && \mbox{correctly classified {\bf Stage 0/1} objects}\nonumber\\
\mbox{33.3\,\%} && \mbox{correctly classified {\bf Stage 2+} objects}\nonumber\\
\mbox{32.1\,\%} && \mbox{misclassified as {\bf Stage 0/1}} \nonumber\\
\mbox{1.4\,\%} && \mbox{misclassified as {\bf Stage 2+}}\nonumber
\end{eqnarray}
\vspace*{-0.7cm}
\begin{figure}
\includegraphics[width=0.5\textwidth]{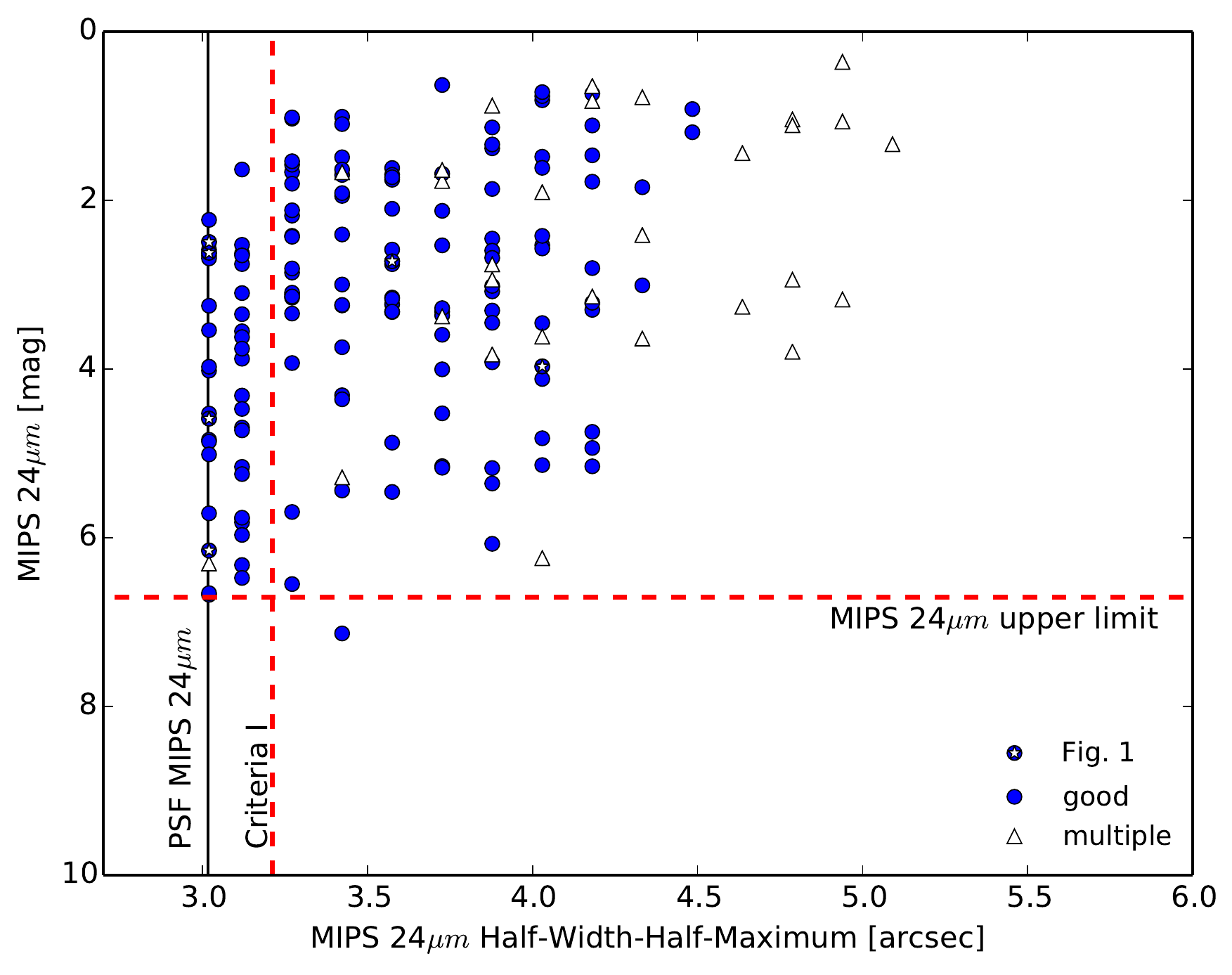}
\caption{\label{sizeobs}Observation counterpart to Figure~\ref{flux}\textcolor{blue}{b}. Measured sizes of sample points of good quality (blue circles) and others which appear to be multiple sources (white triangles) from the \cite{2009:Yusef} sample. The sample sources of bad quality are not included. Points with white stars surrounded by blue circles represent the measurements of the six objects showed in Figure~\ref{Milky Way}.}
\end{figure}
\subsection{Correcting previously estimated SFR}
\label{SFRcorr}
With these criteria it is possible to revise SFR calculations from the literature. Here we investigate the 213 sources classified as {\bf Stage 0/1} by \cite{2009:Yusef}. We use the PSF fitting tool described in Section~\ref{hwsize} with 20 extended models with total HWHMs ranging from 3 to 6\arcsec. We used fewer profiles than for synthetic observations, because for real observations the fitting is done by hand. The measuring tool is robust to distinguish between unresolved and resolved objects. The error increases with increasing size in real observations, but sizes smaller than HWHM $< 3.5$\arcsec \ appear to be robust. We also estimate the total flux of the objects. The values of the magnitudes suffer from errors depending on the background. In Figure~\ref{sizeobs} we plot our measured sizes and magnitudes.

By visually inspecting the MIPS 24\,$\mu$m images, we noted first that about 31.0\,\% of the sources are likely unreliable, of which 56\,\% correspond to parts of diffuse emission erroneously fitted as point sources in the original catalogue and 44\,\% of the sources show evidence of multiplicity.

Further, and more importantly, 32.4\,\% of the sample have HWHM $> 3.5\,\arcsec$ and would clearly be resolved according to our criteria. When we use our original threshold of 3.2$\,\arcsec$, 49.3\,\% of the objects have a larger HWHM, but this may include objects that appear to be larger due to noise, so that this value should be treated more cautiously. This means that at least 63.4\,\% (32.4\,\% + 31.0\,\%) and maybe 80.3\,\% (49.3\% + 31.0\,\%) or more of the \cite{2009:Yusef} sources may therefore not be YSOs, lowering the star formation rate by a factor of three or more and bringing it closer to the value of the SFR estimated from free-free emission (see \citealp{2013:Longmore}).

Given the issues with spurious sources mentioned above, a careful characterization of each source at MIPS 24\,$\mu$m and IRAC 8\,$\mu$m is therefore needed in future to pin down the SFR in the CMZ more accurately.

\section{Discussion}
\label{discuss}
Our ``realistic synthetic observations'' from radiative transfer modelling (see Section~\ref{model}) have shown that more evolved objects (i.\,e.~main-sequence stars) in a low-density ambient density medium could mimic YSOs, for all spectral types discussed in this paper. We found that {\bf Stage 2+} objects with spectral types later than B3, detected in MIPS 24\,$\mu$m, are predicted to have no $5\sigma$-observable counterpart in PACS 70\,$\mu$m, similar to some of the objects classified as YSOs by \cite{2009:Yusef}. 

Unresolved MIPS 24\,$\mu$m sources could be heavily embedded objects ({\bf Stage 0/1})  or more evolved {\bf Stage 2+}, while resolved objects are most likely {\bf Stage 2+} objects. This means that resolved model objects detected in MIPS 24\,$\mu$m in the CMZ are more evolved and therefore likely older than 1 Myr. All these arguments that indicate some objects classified in the CMZ  are not very young embedded YSOs (i.\,e.~less than 1\,Myr) as previously thought. Therefore, the star formation rate of  $0.14\,\mbox{M}_{\odot}\,\mbox{yr}^{-1}$ \citep{2009:Yusef} is likely over-estimated. 

\subsection{Stellar distribution from the IMF}
To get an idea of how many main-sequence stars of various spectral types should be in the CMZ, we performed simple estimations using a \citet{2001:Kroupa} IMF. Assuming a constant SFR of $0.01\,\mbox{M}_{\odot}\,\mbox{yr}^{-1}$ (see \citealp{2013:Longmore}), roughly four O stars and 5183 B stars (including and earlier than B5) should be at the distance of the CMZ. We note, that in this IMF calculation the ages of the stars are equally distributed. Since the lifetime of a MIPS 24\,$\mu$m detectable B star lies, as mass decreases (B0 to B5, $\mbox{M}\in [17.5, 5.9]\,\mbox{M}_\odot$), between 8\,Myr and 113\,Myr, only a few objects should be primordial ({\bf Stage 0/1}, smaller than 1\,Myr), while the majority would be in the main-sequence phase ({\bf Stage 2+}). Not necessarily, all 5138 B stars (including and earlier than B5) could be observed in MIPS 24\,$\mu$m, since sources in lower ambient densities environments can not produce strong MIR emission in order to be detected.

\subsection{Origin of the main-sequence stars}
In the CMZ there are many objects with MIPS 24\,$\mu$m emission west from the Galactic center, which appear not to be part of an active star-forming region analogous to Sgr B2 or Sgr C. Our results show, that some of those may be in fact more evolved objects (i.\,e.~main-sequence stars) and not true YSOs. 

However, more evolved objects with ages of several Myr will not have formed at the observed spot. One could envisage a scenario in which an OB association formed at the current location of Sgr B2, then got disrupted at a later time while orbiting the Galactic center. For example, an object observed at 41.6\,\arcmin\ or 103\,pc from the Galactic center with an average orbital speed of about $140\,\mbox{km}\,\mbox{s}^{-1}$ \citep{2013:Sofue} has an orbital time-scale of 4.5\,Myr, thus all more evolved objects with spectral types earlier than B5 will have completed at least one orbit around the Galactic centre since their birth. 

\subsection{Giants in the CMZ}
It is also possible that giants could mimic YSOs. \cite{2011:An} found that there are supergiants in the CMZ. Evolved objects like supergiants should not produce HII regions. There is not a large overlap with the \cite{2009:Yusef} sample of YSOs but still when recalculating the SFR one should consider these objects, because contamination of the sample would overestimate the SFR. 
We produced synthetic observation of supergiants\footnote{\label{app}The setup and photometric data is provided in the Appendix.} from spectral type B0 to M0 and found that they are distributed amongst resolved sources and could be therefore also be removed by our selection criteria. Further, in the NIR/ MIR color space (e.\,g.~Figure~\ref{flux}\textcolor{blue}{d}) most supergiants are redder than {\bf Stage 0/1} sources but brighter than 12\,mag in $K$ band and 8\,mag in IRAC 8\,$\mu$m, and therefore would be easily distinguishable in color space.

\begin{figure*}
\includegraphics[width=\textwidth]{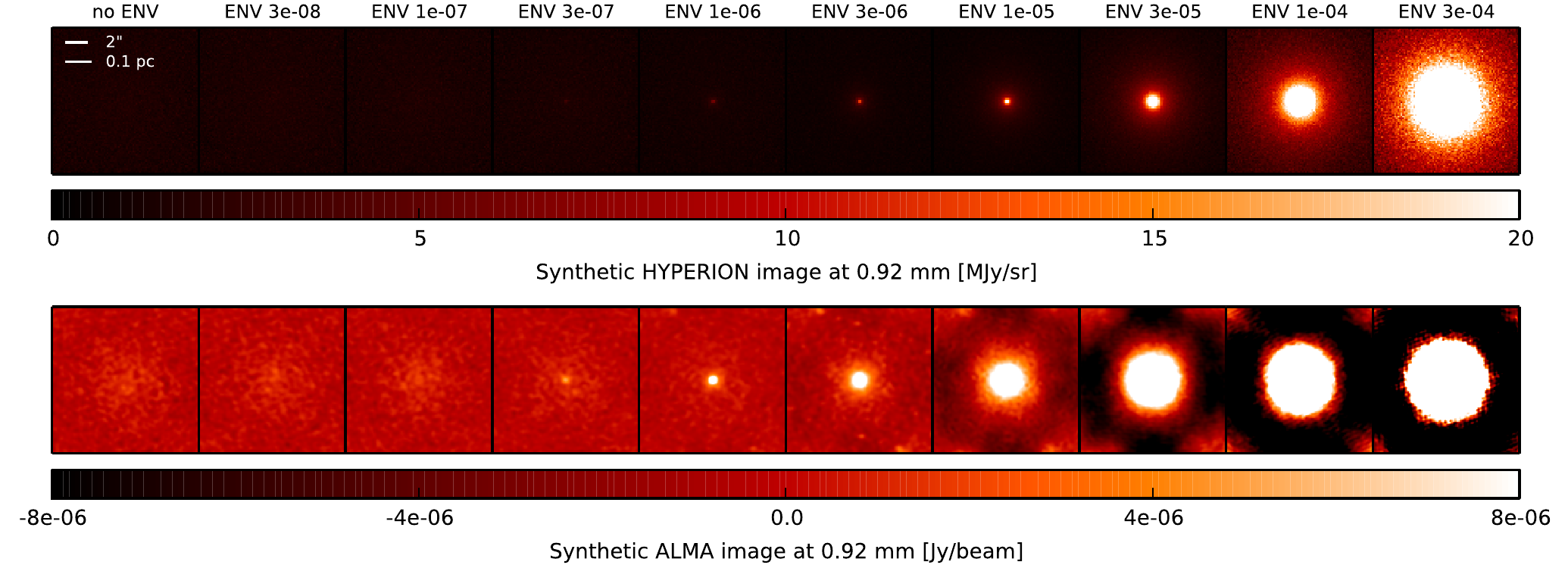}
\caption{\label{mm}Millimeter observation observed through a perfect interferometer of a B5 star (top) and synthetic ALMA observations (bottom). True YSO ({\bf Stage 0/1}) model objects are much brighter and have much more flux on the small scales, while for the most evolved ones, there is very little dense material so most of the emission is extended or not detected. Six models from the left: {\bf Stage 2+}, four models from the right: {\bf Stage 0/1}.}
\end{figure*}

\subsection{Predictions for high-resolution mm observations}
We now briefly describe whether high-resolution mm observations could help us to distinguish between the main-sequence stars ({\bf Stage 2+}) and true YSOs ({\bf Stage 0/1}). In particular, one would expect that the YSOs will show much more dense dust/ gas in the central regions. Millimetre observations of reclassified {\bf Stage 0/1} and {\bf Stage 2+} sources could help us to constrain our predictions, and reduce the number of {\bf Stage 2+} objects misclassified as YSOs. In Figure~\ref{mm}, our 0.9\,mm prediction of ALMA observations using \textsc{Hyperion} and \textsc{Casa} of a B5 star\footnotemark[3] at all evolutionary stages is presented (dust continuum with a bandwidth of 7.5\,GHz, total observing time 1200\,s, beam size $\sim$ 0.5 \arcsec). The four models to the right with the highest infall rate are {\bf Stage 0/1} sources. {\bf Stage 0/1} objects are much brighter and have much more flux on the small scales, while for the most evolved ones, there is very little dense material so most of the emission is extended or not detected. Our analysis of the synthetic ALMA images show that it would be possible to use mm observations to distinguish between the main-sequence stars and the YSOs.

\section{Summary}
\label{summary}
With our realistic synthetic observation from radiative transfer modeling, we have shown that some of the classified YSOs \citep{2009:Yusef} in the CMZ may not necessarily be as young as previously thought (i.\,e.~less then 1\,Myr). In addition, we have shown that some of the observed objects can be better explained by more evolved objects such as main-sequence stars in a constant density interstellar medium. We found that:
\begin{itemize}
\item detection/ non-detection in PACS 70\,$\mu$m is not a reliable handle to distinguish true YSOs from more evolved objects.\vspace*{-0.2cm}
\item resolved, extended objects in MIPS 24\,$\mu$m are unlikely to be deeply embedded YSOs and therefore not truly young.
\end{itemize}
These findings lead us to believe that the SFR in the CMZ estimated by directly counting YSOs \citep{2009:Yusef} is over-estimated by at least a factor of three (and potentially up to a factor of 5). A lower SFR for the CMZ would be in better agreement with estimates from free-free emission (e.\,g.~\citealp{2013:Longmore}). By producing synthetic observations of our YSO models, we have shown that high resolution dust continuum observations with ALMA could in future help to provide a more definite classification of the YSO candidates.

\section*{ACKNOWLEDGEMENT}
We thank the referee for a constructive report that helped us improve the clarity and the strength of the results presented in our paper. This work was partially carried out in the Max Planck Research Group \textit{Star formation throughout the Milky Way Galaxy} at the Max Planck Institute for Astronomy. C. K. is a fellow of the International Max Planck Research School for Astronomy and Cosmic Physics (IMPRS) at the University of Heidelberg, Germany and acknowledges support.

\bibliographystyle{apj}
\bibliography{ref}

\appendix
Here we present input (spectral type, envelope infall rate, stage, ambient medium, dust type) and output (magnitude in UKIDSS 2.2\,$\mu$m, IRAC 8\,$\mu$m and MIPS 24\,$\mu$m and the maximum surface brightness in PACS 70\,$\mu$m, MIPS 24\,$\mu$m HWHM) parameters of the 660 models of YSOs and main-sequence stars in an ambient density environment as well as the models of the supergiants. 
%
%
%

\begin{longtable*}{crcccrrrrc}
\caption*{\ \\\textsc{Parameters and measurements from our models.}}\\
\hline\hline
star & \multicolumn{1}{c}{infall} rate & stage & ambient density & PAH dust & \multicolumn{1}{c}{K 2.2\,$\mu$m} & \multicolumn{1}{c}{IRAC 8\,$\mu$m} & \multicolumn{1}{c}{MIPS 24\,$\mu$m} & \multicolumn{1}{c}{PACS 70\,$\mu$m} & HWHM 24\,$\mu$m \\
 & \multicolumn{1}{c}{[$\mbox{M}_{\odot}\,\mbox{yr}^{-1}$]} &  & [g\,cm$^{-3}$] &  & \multicolumn{1}{c}{[mag]} & \multicolumn{1}{c}{[mag]} & \multicolumn{1}{c}{[mag]} & \multicolumn{1}{c}{[MJy\,sr$^{-1}$]} & [arcsec] \\
 				\hline
 				\footnotesize
 ... & ... & ... & ... & ... & ... & ... & ... & ... & ... \\
B5 & 0 & 2+ & 1$\times 10^{-21}$ & - & 16.29 & 13.21 & 5.46 & 471.53 & 3.11 \\
B5 & 3$\times 10^{-8}$ & 2+ & 1$\times 10^{-21}$ & - & 15.68 & 9.57 & 4.91 & 452.39 & 3.07 \\
B5 & 1$\times 10^{-7}$ & 2+ & 1$\times 10^{-21}$ & - & 14.98 & 8.43 & 4.14 & 407.34 & 3.03 \\
B5 & 3$\times 10^{-7}$ & 2+ & 1$\times 10^{-21}$ & - & 14.04 & 7.46 & 3.31 & 305.25 & 3.02 \\
B5 & 1$\times 10^{-6}$ & 2+ & 1$\times 10^{-21}$ & - & 13.08 & 6.82 & 2.85 & 188.92 & 3.02 \\
B5 & 3$\times 10^{-6}$ & 2+ & 1$\times 10^{-21}$ & - & 12.54 & 6.66 & 2.66 & 220.14 & 3.02 \\
B5 & 1$\times 10^{-5}$ & 0/1 & 1$\times 10^{-21}$ & - & 13.47 & 6.88 & 2.16 & 492.63 & 3.02 \\
B5 & 3$\times 10^{-5}$ & 0/1 & 1$\times 10^{-21}$ & - & 18.81 & 7.51 & 1.97 & 1070.34 & 3.02 \\
B5 & 1$\times 10^{-4}$ & 0/1 & 1$\times 10^{-21}$ & - & 38.45 & 9.73 & 2.84 & 2205.17 & 3.02 \\
B5 & 3$\times 10^{-4}$ & 0/1 & 1$\times 10^{-21}$ & - & 88.48 & 16.45 & 5.99 & 3257.50 & 3.02 \\
 ... & ... & ... & ... & ... & ... & ... & ... & ... & ... \\
				\hline
\end{longtable*}
\footnotesize{The full table is provided in the online material.}


\end{document}